\documentclass[aps,prl,showpacs,floats,twocolumn,superscriptaddress,floatfix,footinbib,longbibliography]{revtex4-1}
\usepackage{graphicx}
\usepackage{amsmath,amsthm,amsfonts,amssymb,bm}
\usepackage{wasysym}
\usepackage[linktocpage,colorlinks=true,linkcolor=blue,citecolor=blue,breaklinks=true,urlcolor=blue]{hyperref}
\usepackage[usenames,dvipsnames]{color}

\usepackage{tikz}
\usepackage{pifont}
\usepackage{xcolor}
\usepackage{amssymb}
\usepackage{oplotsymbl}

\newcommand{\beq}{\begin{equation}}
\newcommand{\eeq}{\end{equation}}
\newcommand{\beqa}{\begin{eqnarray}}
\newcommand{\eeqa}{\end{eqnarray}}
\newcommand{\beqan}{\begin{eqnarray*}}
\newcommand{\eenan}{\end{eqnarray*}}

\definecolor{darkred}{rgb}{0.55, 0.0, 0.0}
\definecolor{orangered}{rgb}{1.0, 0.27, 0.0}
\definecolor{darkgreen}{rgb}{0.0, 0.39, 0.0}
\definecolor{modgreen}{rgb}{0.0, 0.85, 0.0}
\definecolor{limegreen}{rgb}{0.2, 0.8, 0.2}
\definecolor{darkblue}{rgb}{0.0, 0.0, 0.55}
\definecolor{cyan}{rgb}{0.0, 0.8, 0.8}

\begin{document}

\title{Biomimetic Swarm of Active Particles with Coupled Passive-Active Interactions}

\author{Amir Nourhani}
\affiliation{Department of Mechanical Engineering, University of Akron, Akron, Ohio  44325, USA}
\affiliation{Biomimicry Research and Innovation Center, University of Akron, Akron, Ohio  44325, USA}
\affiliation{Departments of Biology, University of Akron, Akron, Ohio  44325, USA}

\begin{abstract}
We study the universal behavior of a class of active colloids whose design is inspired by the collective dynamics of natural systems like schools of fish and flocks of birds. These colloids, with off-center repulsive interaction sites, self-organize into polar swarms exhibiting long-range order and directional motion without significant hydrodynamic interactions. Our simulations show that the system transitions from motile perfect crystals to solid-like, liquid-like and gas-like states depending on noise levels, repulsive interaction strength, and particle density. By analyzing swarm polarity and hexatic bond order parameters, we demonstrate that effective volume fractions based on force-range and torque-range interactions explain the system's universal behavior. This work lays a groundwork for biomimetic applications utilizing the cooperative dynamics of active colloids.
\end{abstract}

\maketitle 

Biological microswimmers and active colloids convert environmental energy into self-propulsion in fluids~\cite{soto2021smart}. 
These entities constitute active suspensions that can self-organize into complex, unpredictable patterns~\cite{Marchetti12,SS2013,koch2011}. Initial studies on dense bacterial suspensions~\cite{Dombrowski04,Cisneros07,Wensink2012,Dunkel13,Gachelin2014} and active biological materials~\cite{Nedelec97,Surrey01,SCDHD2012,SNSTYCO2012,Shelley2016} revealed turbulent-like behavior
with large-scale mixing and significant density and velocity fluctuations due to long-range hydrodynamic interactions.
Artificial particles have been engineered with controlled properties like size, shape, and propulsion mechanisms~\cite{Paxton-JACS-2004, Ebbens10, Wang13-book}. 
In many systems, weak short-range interactions and particle contacts drive collective dynamics, leading to cluster formation or motility-induced phase separation~\cite{howse07,cates2015,Redner2013,Stenhammar2013,Theurkauff2012,Schwarz-Linek2012,Arlt2018,PhysRevLett.123.098001}.
Large-scale directional flow from spontaneous symmetry breaking has been observed in only a few synthetic systems, primarily driven by hydrodynamic interactions~\cite{Quincke:96,Jones:84,Das:13, Bricard13,Bricard15,Karani2019}.

We study the universal behavior of a class of active colloids whose design is inspired by the collective dynamics observed in natural systems such as schools of fish, herds of deer, and flocks of birds. These experimentally realizable colloids~\cite{nourhani2017}, featuring off-center repulsive interaction sites, exhibit long-range order and coherent directional collective motion. Similarly to the animals mentioned above, these microparticles avoid aggregation and spontaneously generate system-scale polar swarms, even with negligible hydrodynamic interactions. This capability positions them as a transformative platform for biomimetic applications, overcoming current barriers in utilizing microswimmer swarms for transport and engineering purposes driven by cooperative behavior.

\begin{figure}[b!]
    \includegraphics[width=0.9\columnwidth]{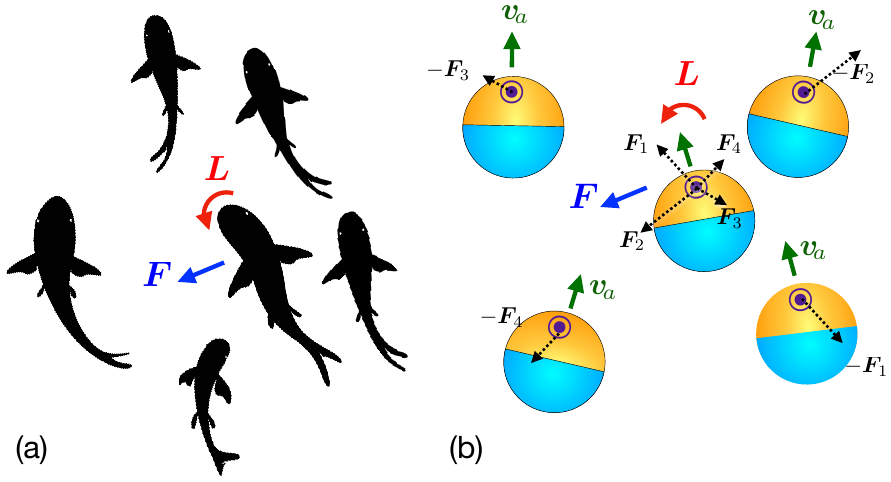}
    \caption{
(a) A fish within a school dynamically avoids contact with others by moving to free space and aligning its direction of motion by exerting torque on itself. (b) Similarly, an active particle with off-center interaction site experiences repulsion from nearby particles; the net force on the off-center site also results in torque about particle's center that align its direction of motion and suppress potential collisions with other particles.
}
    \label{fig:fig1}
\end{figure}

In a simplified phenomenological model of a school of fish, as shown in Fig.~\ref{fig:fig1}(a), animals avoid collisions by moving into free space while aligning with neighbors through effective torques. To replicate this in active colloids, their structure must avoid contact and generate aligning torque. This can be achieved by incorporating additional medium-to-long-range off-center repulsive interaction sites. An experimental example is Janus particles~\cite{nourhani2017} moving in a plane containing magnetic moments oriented perpendicular to the plane of motion, as depicted in Fig.~\ref{fig:fig1}(b). In microswimmers, placing these sites off-center creates {\it passive} torque about the {\it active} particle center, aligning the direction of motion so that when particles approach each other, the torque bends their trajectories apart. The emergent  \mbox{\it coupled} {\it passive-active} interactions lead to polar alignment from the coupling of activity with passive torques, independent of hydrodynamic interactions.

\begin{figure}[t!]
    \includegraphics[width=1\columnwidth]{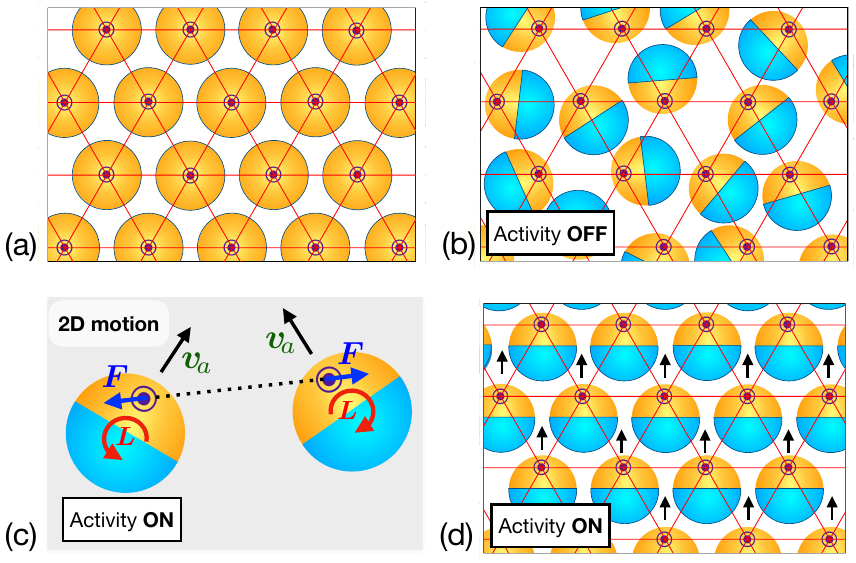}
    \caption{
Top view (a) triangular crystalline structure of passive colloidal spheres with central repulsive interaction sites. (b) In passive colloids with off-center interaction site, the sites arrange on a triangular lattice while the center of the particles will no longer be on a lattice. (c) Two active particles with off-center repulsive interaction sites experience repulsive force while applying torque on each other leading them to turn away from each other and separate. (d) To minimize the forces and torques, the particles move in the same direction like a motile crystal.
}
    \label{fig:fig2}
\end{figure}

To intuitively understand coupled passive-active interactions, consider passive colloidal spheres with strong central repulsive interaction sites in a two-dimensional arrangement. As shown in Fig.~\ref{fig:fig2}(a), these particles repel each other, forming a triangular crystalline structure~\cite{dillmann2013two, ebert2009experimental}. If we design passive particles with off-center interaction sites~\cite{nourhani2017} and place them in the same setup, as shown in Fig.~\ref{fig:fig2}(b), the interaction sites align on a triangular lattice, but the particle centers do not. What happens if these particles are active? As shown in Fig.~\ref{fig:fig2}(c), two isolated active particles repel and exert torques, causing them to turn away and separate due to combined passive torque and active self-propulsion. However, in a dense environment, shown in Fig.~\ref{fig:fig2}(d), they cannot separate and must cooperate to achieve a stable steady state where net interaction force and torque are nearly zero. This steady state occurs when the particles align in the same direction, with centers and off-center sites forming distinct triangular lattices.

In the regime of no noise, active particles with off-center repulsive interaction sites will form motile  perfect crystals, shown in Fig.~\ref{fig:fig2}(d). As noise increases, two new dynamics emerge: translational diffusion of the particle centers and rotational diffusion of the particles' directions of motion. These effects cause the particle centers to deviate from the perfect crystal sites into solid-like, liquid-like and gas-like states depending on noise levels, repulsive interaction strength, and particle density. However, torques still align the particles, allowing the swarm to maintain directional motion, though the arrangement of particle centers may becomes disordered. In the regime of high noise intensity, the coupled passive-active interactions are no longer strong enough to align the particles, resulting in the loss of directional collective motion. In our study, we focus on the dynamics and phases of the polar state, well beyond the regime of dynamics near the isotropic-to-polar transition.~\cite{NourhaniSaintillan2021}
We quantify the collective dynamics using the swarm polarity $|\bm{p}|$, along with the spatial arrangement and degree of crystallinity of the particles, through the local $\smash{\Psi_\text{local}^{(6)}}$ and global $\smash{\Psi_\text{global}^{(6)}}$  hexatic bond orders, as defined below.

The polarity $|\bm{p}|$ of the swarm of $N$ particles is defined by the magnitude of the average direction of active motion
$
\bm{p} = \smash{\langle \hat{\bm n} \rangle} = N^{-1} \sum_{j=1}^N \hat{\bm n}_j
$
where ``director'' $\hat{\bm n}_j$ is the director of the $j$th swimmer's active self-propulsion velocity. We are also interested in the solid-like
and fluid-like behavior as a result of the local and global arrangement of the particles. 
For particle $j$ surrounded by $N_{nn}$ nearest neighbors, determined by Voronoi tessellation, the complex hexatic bond-order parameter 
$
\psi^{(6)}_j = N_{nn}^{-1} \sum_{k=1}^{N_{nn}} e^{i 6 \theta_{jk}}
$
measures how close the nearest-neighbor arrangement around a particle is to a perfect sixfold rotational symmetry. 
The angle $\theta_{jk}$ is the angle between the relative position of the particle $j$ and its $k$th neighbor with respect to the $x$ axis. The value of $|\psi^{(6)}_j|$ varies between 0 and the maxiuum of 1 for perfect hexagonal order in the nearest neighbors.
The local and global orders of crystallinity are characterized by the local bond order $\smash{\Psi_\text{local}^{(6)} = \langle |\psi_j^{(6)}| \rangle}$
and global bond order 
$\smash{\Psi_\text{global}^{(6)} = |\langle \psi_j^{(6)} \rangle|}$,
which differ in the order of taking the modulus, $\smash{|\cdot|}$, and particle average, $\smash{\langle \cdot \rangle:= N^{-1}\sum_{j=1}^N}$.

In our study, each square simulation box of side length $L$ under periodic boundary conditions contains $N$ particles of radius $a$, each with a director $\hat{\bm n} \equiv (\cos\theta, \sin\theta)$, moving with an active self-propulsion velocity $\bm{v}_a = v_a \hat{\bm n}$ in a fluid of viscosity $\mu$, while experiencing translational and orientational diffusion with coefficients
\hbox{$D_\text{t}=k_B T/C_\text{t}$}
and \hbox{$D_\text{o}=k_B T/C_\text{o}$}, respectively. Here,
$C_\text{t} = \frac{3}{4} a^{-2}C_\text{o} = 6\pi \mu a$ is the Stokes drag coefficient.
For each particle, the off-center repulsive interaction site is located at $\frac{3}{4} a \hat{\bm n}$ from the center and the interaction sites are modeled by a force between two magnetic moments perpendicular to the direction of motion~\cite{nourhani2017}. The site in the $j$th particle repels the site of the $k$th particle by a force $\bm{F}_{jk}({\bm r}) = f {\bm r}_{jk}/ r_{jk}^5$, where $f$ defines the strength of the repulsion and ${\bm r}_{jk}$ is the relative position vector from the $j$th to the $k$th site, with $r_{jk} = |{\bm r}_{jk}|$. The total repulsive force acting on the $k$th particle is $\bm{F}_{k} = \sum_{j=1}^N \bm{F}_{jk}$, which leads to an additional translational velocity $C_\text{t}^{-1} \bm{F}_{k}$ and angular velocity $C_\text{o}^{-1} \bm{L}_{k}$, with the torque about the $k$th particle center given by $\bm{L}_{k} = \frac{3}{4} a \hat{\bm n}_k \times \bm{F}_{k}$.

\begin{figure*}[t!]
    \includegraphics[width=0.95\textwidth]{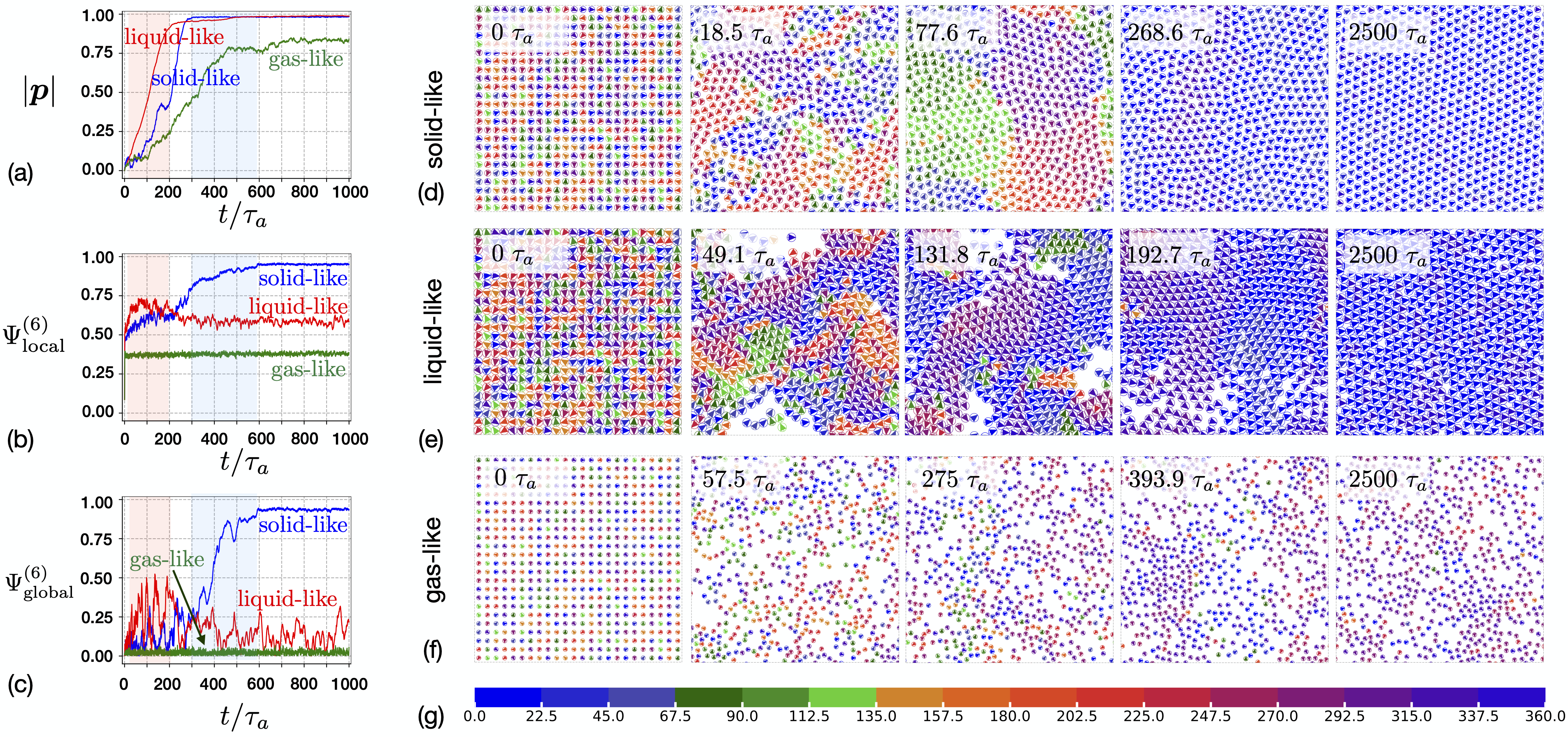}
    \caption{
    The curves for (a) polarization, (b) local bond order, and (c) global bond order as a function of time for systems identified by $(\text{Pe}, \mathcal{F}, \phi)$, exhibiting solid-like $(50, 100, 0.431)$, liquid-like $(500, 1, 0.785)$, and gas-like $(50, 10, 0.196)$ dynamics, with corresponding frames presented in (d), (e), and (f), respectively, starting from a square lattice with random initial directions. (g) The color map for the direction $\theta$ of the swimmer director in degrees. The supplementary video Fig3-video.m4v demonstrates the time evolutions of swarms.
     }
    \label{fig:fig3}
\end{figure*}

In the absence of off-center repulsive interaction sites (\mbox{$f=0$}), the P\'{e}clet number $\text{Pe} = a v_a/D_\text{t}$ is the dimensionless parameter that quantifies the rate of translational displacement relative to spatial diffusion. When off-center repulsion sites are incorporated into the particle structure ($f\neq 0$), two effects emerge. One is the repulsive force that tries to keep the particles apart, tending to place them on a hexagonal lattice (Fig.~\ref{fig:fig2}(d)), while the active propulsive force $C_\text{t} v_a$ may bring the particles closer. The competition between these two forces is quantified by the dimensionless parameter $\mathcal{F} = fa^{-4}/(C_\text{t} v_a)$. The other effect is the competition between the aligning torque about the particle center due to repulsive interactions and the dis-aligning torque $C_\text{o} D_\text{o}$ due to random noise, quantified by $\Omega = (\frac{3}{4}a) fa^{-4}/ (C_\text{o} D_\text{o}) = \frac{3}{4} \mathcal{F} \, \text{Pe}$. This quantity is not an independent dimensionless parameter. Therefore, in our study, we use $\mathcal{F}$ and $\text{Pe}$ as the independent parameters that identify the system.

Therefore, working in units of length $a$ and time $a/v_a$, the governing equations for the position $\bm{x}$ of the particle center and orientation $\theta$ of the particle director $\hat{\bm n}$ are
\begin{align}
\frac{d\bm{x}_k}{dt} 
&= 
\mathcal{F}
\tilde{\bm F}_k
+  
 \sum_{j=1}^N \frac{\bm{r}_{jk}}{r_{jk}} v_{jk}^{(\text{c.c.})}
+ 
\hat{\bm n}_k
+ 
 \sqrt{2\text{Pe}^{-1}} \, \bm{\xi}, \\
\frac{d\theta_k}{d\tilde{t}} 
& = 
\left(\tfrac{3}{4}\right)^2 \mathcal{F}   \, \hat{\bm n}_k \times \tilde{\bm F}_k
+ 
 \sqrt{\tfrac{3}{2}\text{Pe}^{-1}} \, \zeta,
\end{align}
where $\tilde{\bm F}_k = (f a^{-4})^{-1}{\bm F}_k$ is the net dimensionless force on the $i$th off-center site due to other off-center sites. The term $v_{jk}^{(\text{c.c.})} = 10[(r_{jk}-1)^{-2}-1][1-H(r_{jk}-2)]$, with $H(r)$ being the Heaviside step function, represents the speed resulting from the soft-potential approximation of hard-sphere center-to-center repulsion, and $\xi_x$, $\xi_y$, and $\zeta$ are zero-mean unit-strength gaussian white noises.

We simulated for $N = 24^2=576$ particles over a duration of 2500 $\tau_p$, where $\tau_p = a/v_a$ is the time required for an active particle to translate a distance equal to its radius. The simulations covered a range of volume fractions, $\phi = \pi a^2/\ell^2$, from 0.126 ($\ell=5 a$) to 0.785 ($\ell=2 a$), where $\ell = L/\sqrt{N}$ is the average inter-particle distance. The simulations started with particles positioned on a square lattice with a lattice constant $\ell$ and random initial orientations. The reported values are averages at the steady state.

We intuitively define solid-like behavior as a state where particles cannot change their neighbors during collective motion. In liquid-like behavior, particles may change their neighbors slowly, while in gas-like polar dynamics, particles can easily change their neighbors and make large scale displacements. Fig.~\ref{fig:fig3} shows the results for three examples of (Pe, $\mathcal{F}$, $\phi$) demonstrating solid-like $(50, 100, 0.431)$, liquid-like $(100, 1, 0.785)$, and gas-like $(50, 10, 0.196)$ behavior. Figures~\ref{fig:fig3}(a)-(c) show the time evolution of three order parameters: polarity $|\bm{p}|$, local bond order $\Psi_\text{local}^{(6)}$, and global bond order $\Psi_\text{global}^{(6)}$. Figures~\ref{fig:fig3}(d)-(f) show the corresponding snapshots at different times, with the last frame representing an example of the steady state.

\begin{figure*}[t]
    \includegraphics[width=0.95\textwidth]{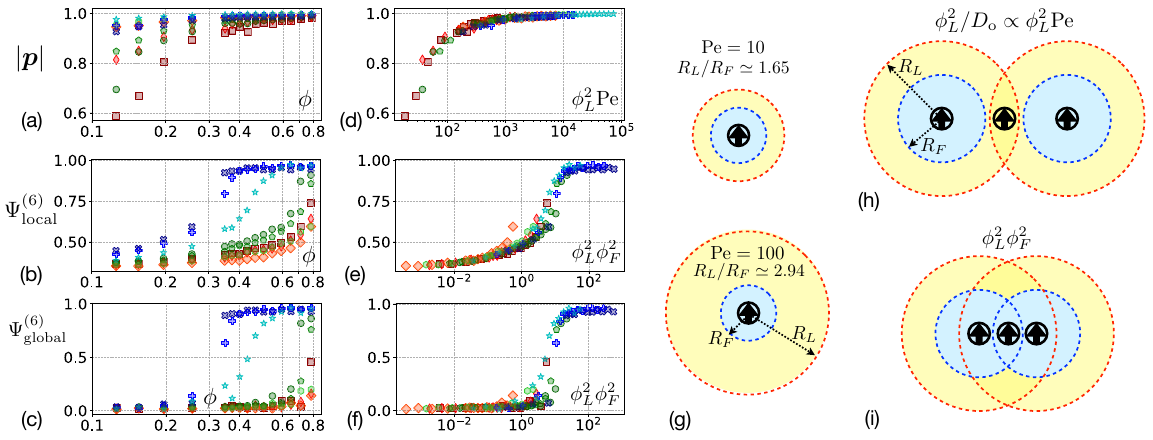}
    \caption{
    (a) Polarization, (b) local bound order, and (c) global bond order as a function of volume fraction $\phi$ for a set of \mbox{(Pe, $\mathcal{F}$)} represented by
    {\color{darkred}  
      $\blacksquare$ 
      }
      (50, 10), 
      {\color{red} 
      $\blacklozenge$
      }
       (100, 5), 
       {\color{orangered}
       $\rhombusfill$ 
       }
       (500, 1),
      {\color{darkgreen} 
      $\circletfill$}
       (50, 20), 
      {\color{modgreen} 
      $\pentagofill$}
       (100, 10), 
       {\color{limegreen}
       $\hexagofill$}
        (500, 2), 
        {\color{darkblue}
        $\mathord{\text{\ding{54}}}$}
         (50, 100), 
         {\color{blue}
         $\mathord{\text{\ding{58}}}$} 
         (100, 50), 
         {\color{cyan}
         $\bigstar$}
          \mbox{(500, 10)}. The master curves for (d) polarization versus $\phi_L^2 \text{P}$ , (e) local bond order versus $\phi_L^2 \phi_F^2$, and (f) global bond order versus $\phi_L^2 \phi_F^2$.
     (g) The ratio $R_L/R_F \propto \text{Pe}^{1/4} > 1$ of torque-range to force-range effective radii. (h) a particle in the overlap region of two torque-range circles. (i) a particle in the overlap region of both two torque-range circles and two force-range circles. 
     }
    \label{fig:fig4}
\end{figure*}

As shown in Fig.~\ref{fig:fig3}(a), all three systems eventually reach a polarized state. The liquid-like system approaches its steady state sooner than the others, as particles can quickly change neighbors and rearrange. In the gas-like system, particles can reorganize more easily, but it takes longer for particles to influence each other's orientation and for the swarm to polarize. In the solid-like behavior, the system is dense enough, and the repulsive interactions are strong enough that the particles begin to reorient locally, creating polarized regions that gradually grow and merge, leading to a fully polarized state.

Figures~\ref{fig:fig3}(b) and (c) show the behavior of the local $\Psi_\text{local}^{(6)}$ and global $\Psi_\text{global}^{(6)}$ bond order parameters, respectively. The solid-like system, after polarization, shows a transient state with hexagonal regions connected by a grain boundary, as seen in Fig.~\ref{fig:fig3}(d). The coexistence of these regions is marked by blue bands in Figs.~\ref{fig:fig3}(a)-(c). During this phase, the local bond order is higher than in the liquid-like and gas-like systems, but the boundary distorts the global order. Eventually, the grain boundaries dissolve, and the system attains global hexagonal order, though slightly distorted by noise. In dense systems or those with strong repulsion, kinetic trapping may prevent this dissolution, blocking the transition to global hexagonal symmetry, even if thermodynamically feasible.

The liquid-like system exhibits a transient state reminiscent of motility-induced phase separation, where particles aggregate and create free volume, as shown in Fig.~\ref{fig:fig3}(e). The local structures approach hexagonal symmetry, leading to an increase in local and global bond order parameters, marked by the red bands in Figs.~\ref{fig:fig3}(a)-(c). However, the particles reorient due to the torques induced by the off-center sites, aligning and repelling each other, which disrupts the local hexagonal symmetry. This results in a reduction of local and global bond order parameters at steady state. In the gas-like system, the local bond order is less than $\tfrac{1}{2}$, and the global bond order is nearly zero, indicating an absence of global hexagonal symmetry.

Figure~\ref{fig:fig4}(a)-(c) shows the polarization, local bond order, and global bond order, respectively, for a set of pairs (Pe, $\mathcal{F}$)
as a function of the volume fraction $\phi$. The trends between polarity and bond order do not appear to be similar. For example,
in the cases of 
{\color{blue} $\mathord{\text{\ding{58}}}$} (100, 50)
 and {\color{cyan} $\bigstar$} \mbox{(500, 10)},
over most of the explored region of $\phi$, the polarization for
{\color{cyan} $\bigstar$} exceeds that of {\color{blue} $\mathord{\text{\ding{58}}}$},
while for the bond orders, that trend does not hold.
As another example, the polarization for 
{\color{darkred}  $\blacksquare$} (50, 10)
 is always less than for 
{\color{orangered} $\rhombusfill$} (500, 1),
while the bond order is higher.
Therefore, to deduce a trend and understand the interplay between the parameters, we introduce the concepts of force-range $\phi_F$ and torque-range $\phi_L$ volume fractions.

We start by defining the force-range $R_F$ and torque-range $R_L$ effective radii, shown in Fig.~\ref{fig:fig4}(g). Consider two swimmers approaching each other until they reach an approximate distance of $2 R_F$, where the self-propulsion force is balanced by the off-center repulsion, keeping them apart. This gives $f(2 R_F)^{-4} \approx C_t v_a$, leading to the force-range effective radius $R_F  \approx (a/2) \mathcal{F}^{1/4}$. Similarly, we define the torque-range effective radius $R_L$ as the distance at which the orientational noise is balanced by the repulsive torque, $(\frac{3}{4}a) f(2 R_L)^{-4} \approx C_\text{o} D_\text{o}$, leading to $R_L  \approx (a/2) \Omega^{1/4}$. Correspondingly, we can define a force-range volume fraction $\phi_F = N \pi R_F^2/L^2 = \frac{1}{4} \phi \mathcal{F}^{1/2}$ and, similarly, a torque-range volume fraction $\phi_L = \frac{1}{4} \phi \Omega^{1/2}$.

The ratio of these effective radii and volume fractions
\begin{equation}
\frac{R_L}{R_F}
=
\left(\frac{\phi_L}{\phi_F}\right)^{1/2}
=
\left(\frac{\Omega}{\mathcal{F}}\right)^{1/4}
=
\left(\frac{3}{4} \,\text{Pe}\right)^{1/4}
\end{equation}
is greater than one in the range of our study (Pe $\geq 50$). As shown in Fig.~\ref{fig:fig4}(g), since $R_L > R_F$, if two particles approach each other, their torque-range effective circles overlap first. Thus, the particles can align and polarize the swarm in a liquid-like or gas-like manner within the parameter domain where their average interparticle distance is less than $2 R_L$ and more than $2 R_F$. Moreover, accounting for overlaps of the effective circles, both ${\phi_L}$ and ${\phi_F}$ can exceed one.

With the definitions of effective volume fractions $\phi_F$ and $\phi_L$, we can construct composite dimensionless quantities to explain the system's universal behavior and collapse the data in Fig.~\ref{fig:fig4}(a)-(c) into master curves. Starting with polarization, as shown in Fig.~\ref{fig:fig4}(h), a particle must fall within the overlap of the torque-range circles of at least two others to balance orientational noise, making the interactions a function of $\phi_L^2$. Swarm polarization also depends on the average orientation of the particles, which is influenced by the amplitude of the oscillation of particle orientation around the swarm's polarization direction. Lower orientational diffusivity reduces this oscillation amplitude, introducing a factor of $D_\text{o}^{-1} \propto \text{Pe}$. As shown in Fig.~\ref{fig:fig4}(d), plotting polarization against the composite dimensionless parameter $\phi_L^2 \text{Pe} \propto \phi_L^2/ D_\text{o}$ reveals a master curve, with polarization increasing as this parameter grows.

For bond orders, we have two objectives. As shown in Fig.~\ref{fig:fig4}(i), one goal is to keep the particle aligned with other particles, which requires them to fall within the overlap of the torque-range circles of two particles, introducing a factor of $\phi_L^2$. The other goal is to ensure the particles stay apart, which requires them to be within the overlap of the force-range circles, quantified by $\phi_F^2$. As shown in Figs.~\ref{fig:fig4}(e) and \ref{fig:fig4}(f), plotting local and global bond orders against $\phi_F^2 \phi_L^2$ results in the data collapsing into master curves.

In conclusion, this study discusses a paradigm for engineering collective dynamics in active colloids, specifically through the design of off-center repulsive interaction sites. By mimicking the behavior of natural systems like schools of fish, these colloids demonstrate the ability to self-organize into polar swarms with long-range order and coherent directional motion, even without significant hydrodynamic interactions. The findings highlight the emergence of coupled passive-active interactions that enable unique control over particle alignment and motion. This work not only advances our understanding of active matter but also provides a versatile platform for developing biomimetic applications in transport, engineering, and microrobotics. The study paves the way for future exploration of collective behaviors in active systems, particularly in overcoming the challenges of aggregation and achieving controlled, coherent motion.

I extend my sincere gratitude to Paul E. Lammert and Vincent H. Crespi for their insightful discussions and valuable comments. I also thank Seyed Amin Nabavizadeh for help with GPU programming and Rafe Md. Abu Zayed for running some simulations. I wishes to acknowledge the support from the National Science Foundation CAREER award, grant number CBET-2238915.


\begin{thebibliography}{35}%
\makeatletter
\providecommand \@ifxundefined [1]{%
 \@ifx{#1\undefined}
}%
\providecommand \@ifnum [1]{%
 \ifnum #1\expandafter \@firstoftwo
 \else \expandafter \@secondoftwo
 \fi
}%
\providecommand \@ifx [1]{%
 \ifx #1\expandafter \@firstoftwo
 \else \expandafter \@secondoftwo
 \fi
}%
\providecommand \natexlab [1]{#1}%
\providecommand \enquote  [1]{``#1''}%
\providecommand \bibnamefont  [1]{#1}%
\providecommand \bibfnamefont [1]{#1}%
\providecommand \citenamefont [1]{#1}%
\providecommand \href@noop [0]{\@secondoftwo}%
\providecommand \href [0]{\begingroup \@sanitize@url \@href}%
\providecommand \@href[1]{\@@startlink{#1}\@@href}%
\providecommand \@@href[1]{\endgroup#1\@@endlink}%
\providecommand \@sanitize@url [0]{\catcode `\\12\catcode `\$12\catcode
  `\&12\catcode `\#12\catcode `\^12\catcode `\_12\catcode `\%12\relax}%
\providecommand \@@startlink[1]{}%
\providecommand \@@endlink[0]{}%
\providecommand \url  [0]{\begingroup\@sanitize@url \@url }%
\providecommand \@url [1]{\endgroup\@href {#1}{\urlprefix }}%
\providecommand \urlprefix  [0]{URL }%
\providecommand \Eprint [0]{\href }%
\providecommand \doibase [0]{http://dx.doi.org/}%
\providecommand \selectlanguage [0]{\@gobble}%
\providecommand \bibinfo  [0]{\@secondoftwo}%
\providecommand \bibfield  [0]{\@secondoftwo}%
\providecommand \translation [1]{[#1]}%
\providecommand \BibitemOpen [0]{}%
\providecommand \bibitemStop [0]{}%
\providecommand \bibitemNoStop [0]{.\EOS\space}%
\providecommand \EOS [0]{\spacefactor3000\relax}%
\providecommand \BibitemShut  [1]{\csname bibitem#1\endcsname}%
\let\auto@bib@innerbib\@empty
\bibitem [{\citenamefont {Soto}\ \emph {et~al.}(2022)\citenamefont {Soto},
  \citenamefont {Karshalev}, \citenamefont {Zhang}, \citenamefont {de~Avila},
  \citenamefont {Nourhani},\ and\ \citenamefont {Wang}}]{soto2021smart}%
  \BibitemOpen
  \bibfield  {author} {\bibinfo {author} {\bibfnamefont {F.}~\bibnamefont
  {Soto}}, \bibinfo {author} {\bibfnamefont {E.}~\bibnamefont {Karshalev}},
  \bibinfo {author} {\bibfnamefont {F.}~\bibnamefont {Zhang}}, \bibinfo
  {author} {\bibfnamefont {B.~E.~F.}\ \bibnamefont {de~Avila}}, \bibinfo
  {author} {\bibfnamefont {A.}~\bibnamefont {Nourhani}}, \ and\ \bibinfo
  {author} {\bibfnamefont {J.}~\bibnamefont {Wang}},\ }\href@noop {} {\bibfield
   {journal} {\bibinfo  {journal} {Chem. Rev.}\ }\textbf {\bibinfo {volume}
  {122}},\ \bibinfo {pages} {5365} (\bibinfo {year} {2022})}\BibitemShut
  {NoStop}%
\bibitem [{\citenamefont {Marchetti}\ \emph {et~al.}(2013)\citenamefont
  {Marchetti}, \citenamefont {Joanny}, \citenamefont {Ramaswamy}, \citenamefont
  {Liverpool}, \citenamefont {Prost}, \citenamefont {Rao},\ and\ \citenamefont
  {{Aditi Simha}}}]{Marchetti12}%
  \BibitemOpen
  \bibfield  {author} {\bibinfo {author} {\bibfnamefont {M.~C.}\ \bibnamefont
  {Marchetti}}, \bibinfo {author} {\bibfnamefont {J.~F.}\ \bibnamefont
  {Joanny}}, \bibinfo {author} {\bibfnamefont {S.}~\bibnamefont {Ramaswamy}},
  \bibinfo {author} {\bibfnamefont {T.~B.}\ \bibnamefont {Liverpool}}, \bibinfo
  {author} {\bibfnamefont {J.}~\bibnamefont {Prost}}, \bibinfo {author}
  {\bibfnamefont {M.}~\bibnamefont {Rao}}, \ and\ \bibinfo {author}
  {\bibfnamefont {R.}~\bibnamefont {{Aditi Simha}}},\ }\href@noop {} {\bibfield
   {journal} {\bibinfo  {journal} {Rev. Mod. Phys.}\ }\textbf {\bibinfo
  {volume} {85}},\ \bibinfo {pages} {1143} (\bibinfo {year}
  {2013})}\BibitemShut {NoStop}%
\bibitem [{\citenamefont {Saintillan}\ and\ \citenamefont
  {Shelley}(2013)}]{SS2013}%
  \BibitemOpen
  \bibfield  {author} {\bibinfo {author} {\bibfnamefont {D.}~\bibnamefont
  {Saintillan}}\ and\ \bibinfo {author} {\bibfnamefont {M.~J.}\ \bibnamefont
  {Shelley}},\ }\href@noop {} {\bibfield  {journal} {\bibinfo  {journal}
  {Comptes Rendus Physique}\ }\textbf {\bibinfo {volume} {14}},\ \bibinfo
  {pages} {497} (\bibinfo {year} {2013})}\BibitemShut {NoStop}%
\bibitem [{\citenamefont {Koch}\ and\ \citenamefont
  {Subramanian}(2011)}]{koch2011}%
  \BibitemOpen
  \bibfield  {author} {\bibinfo {author} {\bibfnamefont {D.~L.}\ \bibnamefont
  {Koch}}\ and\ \bibinfo {author} {\bibfnamefont {G.}~\bibnamefont
  {Subramanian}},\ }\href@noop {} {\bibfield  {journal} {\bibinfo  {journal}
  {Annu. Rev. Fluid Mech.}\ }\textbf {\bibinfo {volume} {43}},\ \bibinfo
  {pages} {637} (\bibinfo {year} {2011})}\BibitemShut {NoStop}%
\bibitem [{\citenamefont {Dombrowski}\ \emph {et~al.}(2004)\citenamefont
  {Dombrowski}, \citenamefont {Cisneros}, \citenamefont {Chatkaew},
  \citenamefont {Goldstein},\ and\ \citenamefont {Kessler}}]{Dombrowski04}%
  \BibitemOpen
  \bibfield  {author} {\bibinfo {author} {\bibfnamefont {C.}~\bibnamefont
  {Dombrowski}}, \bibinfo {author} {\bibfnamefont {L.}~\bibnamefont
  {Cisneros}}, \bibinfo {author} {\bibfnamefont {S.}~\bibnamefont {Chatkaew}},
  \bibinfo {author} {\bibfnamefont {R.~E.}\ \bibnamefont {Goldstein}}, \ and\
  \bibinfo {author} {\bibfnamefont {J.~O.}\ \bibnamefont {Kessler}},\
  }\href@noop {} {\bibfield  {journal} {\bibinfo  {journal} {Phys. Rev. Lett.}\
  }\textbf {\bibinfo {volume} {93}},\ \bibinfo {pages} {098103} (\bibinfo
  {year} {2004})}\BibitemShut {NoStop}%
\bibitem [{\citenamefont {Cisneros}\ \emph {et~al.}(2007)\citenamefont
  {Cisneros}, \citenamefont {Cortez}, \citenamefont {Dombrowski}, \citenamefont
  {Goldstein},\ and\ \citenamefont {Kessler}}]{Cisneros07}%
  \BibitemOpen
  \bibfield  {author} {\bibinfo {author} {\bibfnamefont {L.~H.}\ \bibnamefont
  {Cisneros}}, \bibinfo {author} {\bibfnamefont {R.}~\bibnamefont {Cortez}},
  \bibinfo {author} {\bibfnamefont {C.}~\bibnamefont {Dombrowski}}, \bibinfo
  {author} {\bibfnamefont {R.~E.}\ \bibnamefont {Goldstein}}, \ and\ \bibinfo
  {author} {\bibfnamefont {J.~O.}\ \bibnamefont {Kessler}},\ }\href@noop {}
  {\bibfield  {journal} {\bibinfo  {journal} {Exp. Fluids}\ }\textbf {\bibinfo
  {volume} {43}},\ \bibinfo {pages} {737} (\bibinfo {year} {2007})}\BibitemShut
  {NoStop}%
\bibitem [{\citenamefont {Wensink}\ \emph {et~al.}(2012)\citenamefont
  {Wensink}, \citenamefont {Dunkel}, \citenamefont {Heidenreich}, \citenamefont
  {Drescher}, \citenamefont {Goldstein}, \citenamefont {L{\"o}wen},\ and\
  \citenamefont {Yeomans}}]{Wensink2012}%
  \BibitemOpen
  \bibfield  {author} {\bibinfo {author} {\bibfnamefont {H.~H.}\ \bibnamefont
  {Wensink}}, \bibinfo {author} {\bibfnamefont {J.}~\bibnamefont {Dunkel}},
  \bibinfo {author} {\bibfnamefont {S.}~\bibnamefont {Heidenreich}}, \bibinfo
  {author} {\bibfnamefont {K.}~\bibnamefont {Drescher}}, \bibinfo {author}
  {\bibfnamefont {R.~E.}\ \bibnamefont {Goldstein}}, \bibinfo {author}
  {\bibfnamefont {H.}~\bibnamefont {L{\"o}wen}}, \ and\ \bibinfo {author}
  {\bibfnamefont {J.~M.}\ \bibnamefont {Yeomans}},\ }\href@noop {} {\bibfield
  {journal} {\bibinfo  {journal} {Proc. Natl. Acad. Sci.}\ }\textbf {\bibinfo
  {volume} {109}},\ \bibinfo {pages} {14308} (\bibinfo {year}
  {2012})}\BibitemShut {NoStop}%
\bibitem [{\citenamefont {Dunkel}\ \emph {et~al.}(2013)\citenamefont {Dunkel},
  \citenamefont {Heidenreich}, \citenamefont {Drescher}, \citenamefont
  {Wensink}, \citenamefont {Bar},\ and\ \citenamefont {Goldstein}}]{Dunkel13}%
  \BibitemOpen
  \bibfield  {author} {\bibinfo {author} {\bibfnamefont {J.}~\bibnamefont
  {Dunkel}}, \bibinfo {author} {\bibfnamefont {S.}~\bibnamefont {Heidenreich}},
  \bibinfo {author} {\bibfnamefont {K.}~\bibnamefont {Drescher}}, \bibinfo
  {author} {\bibfnamefont {H.~H.}\ \bibnamefont {Wensink}}, \bibinfo {author}
  {\bibfnamefont {M.}~\bibnamefont {Bar}}, \ and\ \bibinfo {author}
  {\bibfnamefont {R.~E.}\ \bibnamefont {Goldstein}},\ }\href@noop {} {\bibfield
   {journal} {\bibinfo  {journal} {Phys. Rev. Lett.}\ }\textbf {\bibinfo
  {volume} {110}},\ \bibinfo {pages} {228102} (\bibinfo {year}
  {2013})}\BibitemShut {NoStop}%
\bibitem [{\citenamefont {Gachelin}\ \emph {et~al.}(2014)\citenamefont
  {Gachelin}, \citenamefont {Rousselet}, \citenamefont {Lindner},\ and\
  \citenamefont {Clement}}]{Gachelin2014}%
  \BibitemOpen
  \bibfield  {author} {\bibinfo {author} {\bibfnamefont {J.}~\bibnamefont
  {Gachelin}}, \bibinfo {author} {\bibfnamefont {A.}~\bibnamefont {Rousselet}},
  \bibinfo {author} {\bibfnamefont {A.}~\bibnamefont {Lindner}}, \ and\
  \bibinfo {author} {\bibfnamefont {E.}~\bibnamefont {Clement}},\ }\href@noop
  {} {\bibfield  {journal} {\bibinfo  {journal} {N. J. Phys.}\ }\textbf
  {\bibinfo {volume} {16}},\ \bibinfo {pages} {025003} (\bibinfo {year}
  {2014})}\BibitemShut {NoStop}%
\bibitem [{\citenamefont {Nedelec}\ \emph {et~al.}(1997)\citenamefont
  {Nedelec}, \citenamefont {Surrey}, \citenamefont {Maggs},\ and\ \citenamefont
  {Leibler}}]{Nedelec97}%
  \BibitemOpen
  \bibfield  {author} {\bibinfo {author} {\bibfnamefont {F.~J.}\ \bibnamefont
  {Nedelec}}, \bibinfo {author} {\bibfnamefont {T.}~\bibnamefont {Surrey}},
  \bibinfo {author} {\bibfnamefont {A.~C.}\ \bibnamefont {Maggs}}, \ and\
  \bibinfo {author} {\bibfnamefont {S.}~\bibnamefont {Leibler}},\ }\href@noop
  {} {\bibfield  {journal} {\bibinfo  {journal} {Nature}\ }\textbf {\bibinfo
  {volume} {389}},\ \bibinfo {pages} {305} (\bibinfo {year}
  {1997})}\BibitemShut {NoStop}%
\bibitem [{\citenamefont {Surrey}\ \emph {et~al.}(2001)\citenamefont {Surrey},
  \citenamefont {N\'ed\'elec}, \citenamefont {Leibler},\ and\ \citenamefont
  {Karsenti}}]{Surrey01}%
  \BibitemOpen
  \bibfield  {author} {\bibinfo {author} {\bibfnamefont {T.}~\bibnamefont
  {Surrey}}, \bibinfo {author} {\bibfnamefont {F.}~\bibnamefont {N\'ed\'elec}},
  \bibinfo {author} {\bibfnamefont {S.}~\bibnamefont {Leibler}}, \ and\
  \bibinfo {author} {\bibfnamefont {E.}~\bibnamefont {Karsenti}},\ }\href@noop
  {} {\bibfield  {journal} {\bibinfo  {journal} {Science}\ }\textbf {\bibinfo
  {volume} {292}},\ \bibinfo {pages} {1167} (\bibinfo {year}
  {2001})}\BibitemShut {NoStop}%
\bibitem [{\citenamefont {Sanchez}\ \emph {et~al.}(2012)\citenamefont
  {Sanchez}, \citenamefont {Chen}, \citenamefont {DeCamp}, \citenamefont
  {Heymann},\ and\ \citenamefont {Dogic}}]{SCDHD2012}%
  \BibitemOpen
  \bibfield  {author} {\bibinfo {author} {\bibfnamefont {T.}~\bibnamefont
  {Sanchez}}, \bibinfo {author} {\bibfnamefont {D.}~\bibnamefont {Chen}},
  \bibinfo {author} {\bibfnamefont {S.}~\bibnamefont {DeCamp}}, \bibinfo
  {author} {\bibfnamefont {M.}~\bibnamefont {Heymann}}, \ and\ \bibinfo
  {author} {\bibfnamefont {Z.}~\bibnamefont {Dogic}},\ }\href@noop {}
  {\bibfield  {journal} {\bibinfo  {journal} {Nature}\ }\textbf {\bibinfo
  {volume} {491}},\ \bibinfo {pages} {431} (\bibinfo {year}
  {2012})}\BibitemShut {NoStop}%
\bibitem [{\citenamefont {Sumino}\ \emph {et~al.}(2012)\citenamefont {Sumino},
  \citenamefont {Nagai}, \citenamefont {Shitaka}, \citenamefont {Tanaka},
  \citenamefont {Yoshikawa}, \citenamefont {Chate},\ and\ \citenamefont
  {Oiwa}}]{SNSTYCO2012}%
  \BibitemOpen
  \bibfield  {author} {\bibinfo {author} {\bibfnamefont {Y.}~\bibnamefont
  {Sumino}}, \bibinfo {author} {\bibfnamefont {K.}~\bibnamefont {Nagai}},
  \bibinfo {author} {\bibfnamefont {Y.}~\bibnamefont {Shitaka}}, \bibinfo
  {author} {\bibfnamefont {D.}~\bibnamefont {Tanaka}}, \bibinfo {author}
  {\bibfnamefont {K.}~\bibnamefont {Yoshikawa}}, \bibinfo {author}
  {\bibfnamefont {H.}~\bibnamefont {Chate}}, \ and\ \bibinfo {author}
  {\bibfnamefont {K.}~\bibnamefont {Oiwa}},\ }\href@noop {} {\bibfield
  {journal} {\bibinfo  {journal} {Nature}\ }\textbf {\bibinfo {volume} {483}},\
  \bibinfo {pages} {228} (\bibinfo {year} {2012})}\BibitemShut {NoStop}%
\bibitem [{\citenamefont {Shelley}(2016)}]{Shelley2016}%
  \BibitemOpen
  \bibfield  {author} {\bibinfo {author} {\bibfnamefont {M.~J.}\ \bibnamefont
  {Shelley}},\ }\href@noop {} {\bibfield  {journal} {\bibinfo  {journal} {Annu.
  Rev. Fluid Mech.}\ }\textbf {\bibinfo {volume} {48}},\ \bibinfo {pages} {487}
  (\bibinfo {year} {2016})}\BibitemShut {NoStop}%
\bibitem [{\citenamefont {Paxton}\ \emph {et~al.}(2004)\citenamefont {Paxton},
  \citenamefont {Kistler}, \citenamefont {Olmeda}, \citenamefont {Sen},
  \citenamefont {Angelo}, \citenamefont {Cao}, \citenamefont {Mallouk},
  \citenamefont {Lammert},\ and\ \citenamefont {Crespi}}]{Paxton-JACS-2004}%
  \BibitemOpen
  \bibfield  {author} {\bibinfo {author} {\bibfnamefont {W.~F.}\ \bibnamefont
  {Paxton}}, \bibinfo {author} {\bibfnamefont {K.~C.}\ \bibnamefont {Kistler}},
  \bibinfo {author} {\bibfnamefont {C.~C.}\ \bibnamefont {Olmeda}}, \bibinfo
  {author} {\bibfnamefont {A.}~\bibnamefont {Sen}}, \bibinfo {author}
  {\bibfnamefont {S.~K.~S.}\ \bibnamefont {Angelo}}, \bibinfo {author}
  {\bibfnamefont {Y.~Y.}\ \bibnamefont {Cao}}, \bibinfo {author} {\bibfnamefont
  {T.~E.}\ \bibnamefont {Mallouk}}, \bibinfo {author} {\bibfnamefont {P.~E.}\
  \bibnamefont {Lammert}}, \ and\ \bibinfo {author} {\bibfnamefont {V.~H.}\
  \bibnamefont {Crespi}},\ }\href@noop {} {\bibfield  {journal} {\bibinfo
  {journal} {J. Am. Chem. Soc.}\ }\textbf {\bibinfo {volume} {126}},\ \bibinfo
  {pages} {13424} (\bibinfo {year} {2004})}\BibitemShut {NoStop}%
\bibitem [{\citenamefont {Ebbens}\ and\ \citenamefont
  {Howse}(2010)}]{Ebbens10}%
  \BibitemOpen
  \bibfield  {author} {\bibinfo {author} {\bibfnamefont {S.~J.}\ \bibnamefont
  {Ebbens}}\ and\ \bibinfo {author} {\bibfnamefont {J.~R.}\ \bibnamefont
  {Howse}},\ }\href@noop {} {\bibfield  {journal} {\bibinfo  {journal} {Soft
  Matter}\ }\textbf {\bibinfo {volume} {6}},\ \bibinfo {pages} {726} (\bibinfo
  {year} {2010})}\BibitemShut {NoStop}%
\bibitem [{\citenamefont {Wang}(2013)}]{Wang13-book}%
  \BibitemOpen
  \bibfield  {author} {\bibinfo {author} {\bibfnamefont {J.}~\bibnamefont
  {Wang}},\ }\href@noop {} {\emph {\bibinfo {title} {Nanomachines: Fundamentals
  and Applications}}}\ (\bibinfo  {publisher} {Wiley},\ \bibinfo {year}
  {2013})\BibitemShut {NoStop}%
\bibitem [{\citenamefont {Howse}\ \emph {et~al.}(2007)\citenamefont {Howse},
  \citenamefont {Jones}, \citenamefont {Ryan}, \citenamefont {Gough},
  \citenamefont {Vafabakhsh},\ and\ \citenamefont {Golestanian}}]{howse07}%
  \BibitemOpen
  \bibfield  {author} {\bibinfo {author} {\bibfnamefont {J.~R.}\ \bibnamefont
  {Howse}}, \bibinfo {author} {\bibfnamefont {R.~A.~L.}\ \bibnamefont {Jones}},
  \bibinfo {author} {\bibfnamefont {A.~J.}\ \bibnamefont {Ryan}}, \bibinfo
  {author} {\bibfnamefont {T.}~\bibnamefont {Gough}}, \bibinfo {author}
  {\bibfnamefont {R.}~\bibnamefont {Vafabakhsh}}, \ and\ \bibinfo {author}
  {\bibfnamefont {R.}~\bibnamefont {Golestanian}},\ }\href@noop {} {\bibfield
  {journal} {\bibinfo  {journal} {Phys. Rev. Lett.}\ }\textbf {\bibinfo
  {volume} {99}},\ \bibinfo {pages} {048102} (\bibinfo {year}
  {2007})}\BibitemShut {NoStop}%
\bibitem [{\citenamefont {Cates}\ and\ \citenamefont
  {Tailleur}(2015)}]{cates2015}%
  \BibitemOpen
  \bibfield  {author} {\bibinfo {author} {\bibfnamefont {M.~E.}\ \bibnamefont
  {Cates}}\ and\ \bibinfo {author} {\bibfnamefont {J.}~\bibnamefont
  {Tailleur}},\ }\href@noop {} {\bibfield  {journal} {\bibinfo  {journal}
  {Annu. Rev. Condens. Matt. Phys.}\ }\textbf {\bibinfo {volume} {6}},\
  \bibinfo {pages} {219} (\bibinfo {year} {2015})}\BibitemShut {NoStop}%
\bibitem [{\citenamefont {Redner}\ \emph {et~al.}(2013)\citenamefont {Redner},
  \citenamefont {Hagan},\ and\ \citenamefont {Baskaran}}]{Redner2013}%
  \BibitemOpen
  \bibfield  {author} {\bibinfo {author} {\bibfnamefont {G.~S.}\ \bibnamefont
  {Redner}}, \bibinfo {author} {\bibfnamefont {M.~F.}\ \bibnamefont {Hagan}}, \
  and\ \bibinfo {author} {\bibfnamefont {A.}~\bibnamefont {Baskaran}},\
  }\href@noop {} {\bibfield  {journal} {\bibinfo  {journal} {Phys. Rev. Lett.}\
  }\textbf {\bibinfo {volume} {110}},\ \bibinfo {pages} {055701} (\bibinfo
  {year} {2013})}\BibitemShut {NoStop}%
\bibitem [{\citenamefont {Stenhammar}\ \emph {et~al.}(2013)\citenamefont
  {Stenhammar}, \citenamefont {Tiribocchi}, \citenamefont {Allen},
  \citenamefont {Marenduzzo},\ and\ \citenamefont {Cates}}]{Stenhammar2013}%
  \BibitemOpen
  \bibfield  {author} {\bibinfo {author} {\bibfnamefont {J.}~\bibnamefont
  {Stenhammar}}, \bibinfo {author} {\bibfnamefont {A.}~\bibnamefont
  {Tiribocchi}}, \bibinfo {author} {\bibfnamefont {R.~J.}\ \bibnamefont
  {Allen}}, \bibinfo {author} {\bibfnamefont {D.}~\bibnamefont {Marenduzzo}}, \
  and\ \bibinfo {author} {\bibfnamefont {M.~E.}\ \bibnamefont {Cates}},\ } {\bibfield
  {journal} {\bibinfo  {journal} {Phys. Rev. Lett.}\ }\textbf {\bibinfo
  {volume} {111}},\ \bibinfo {pages} {145702} (\bibinfo {year}
  {2013})}\BibitemShut {NoStop}%
\bibitem [{\citenamefont {Theurkauff}\ \emph {et~al.}(2012)\citenamefont
  {Theurkauff}, \citenamefont {Cottin-Bizonne}, \citenamefont {Palacci},
  \citenamefont {Ybert},\ and\ \citenamefont {Bocquet}}]{Theurkauff2012}%
  \BibitemOpen
  \bibfield  {author} {\bibinfo {author} {\bibfnamefont {I.}~\bibnamefont
  {Theurkauff}}, \bibinfo {author} {\bibfnamefont {C.}~\bibnamefont
  {Cottin-Bizonne}}, \bibinfo {author} {\bibfnamefont {J.}~\bibnamefont
  {Palacci}}, \bibinfo {author} {\bibfnamefont {C.}~\bibnamefont {Ybert}}, \
  and\ \bibinfo {author} {\bibfnamefont {L.}~\bibnamefont {Bocquet}},\
  }\href@noop {} {\bibfield  {journal} {\bibinfo  {journal} {Phys. Rev. Lett.}\
  }\textbf {\bibinfo {volume} {108}},\ \bibinfo {pages} {268303} (\bibinfo
  {year} {2012})}\BibitemShut {NoStop}%
\bibitem [{\citenamefont {Schwarz-Linek}\ \emph {et~al.}(2012)\citenamefont
  {Schwarz-Linek}, \citenamefont {Valeriani}, \citenamefont {Cacciuto},
  \citenamefont {Cates}, \citenamefont {Marenduzzo}, \citenamefont {Morozov},\
  and\ \citenamefont {Poon}}]{Schwarz-Linek2012}%
  \BibitemOpen
  \bibfield  {author} {\bibinfo {author} {\bibfnamefont {J.}~\bibnamefont
  {Schwarz-Linek}}, \bibinfo {author} {\bibfnamefont {C.}~\bibnamefont
  {Valeriani}}, \bibinfo {author} {\bibfnamefont {A.}~\bibnamefont {Cacciuto}},
  \bibinfo {author} {\bibfnamefont {M.~E.}\ \bibnamefont {Cates}}, \bibinfo
  {author} {\bibfnamefont {D.}~\bibnamefont {Marenduzzo}}, \bibinfo {author}
  {\bibfnamefont {A.~N.}\ \bibnamefont {Morozov}}, \ and\ \bibinfo {author}
  {\bibfnamefont {W.~C.~K.}\ \bibnamefont {Poon}},\ }{\bibfield  {journal} {\bibinfo  {journal} {Proc.
  Natl. Acad. Sci.}\ }\textbf {\bibinfo {volume} {109}},\ \bibinfo {pages}
  {4052} (\bibinfo {year} {2012})}\BibitemShut {NoStop}%
\bibitem [{\citenamefont {Arlt}\ \emph {et~al.}(2018)\citenamefont {Arlt},
  \citenamefont {Martinez}, \citenamefont {Dawson}, \citenamefont {Pilizota},\
  and\ \citenamefont {Poon}}]{Arlt2018}%
  \BibitemOpen
  \bibfield  {author} {\bibinfo {author} {\bibfnamefont {J.}~\bibnamefont
  {Arlt}}, \bibinfo {author} {\bibfnamefont {V.~A.}\ \bibnamefont {Martinez}},
  \bibinfo {author} {\bibfnamefont {A.}~\bibnamefont {Dawson}}, \bibinfo
  {author} {\bibfnamefont {T.}~\bibnamefont {Pilizota}}, \ and\ \bibinfo
  {author} {\bibfnamefont {W.~C.~K.}\ \bibnamefont {Poon}},\ }\href@noop {}
  {\bibfield  {journal} {\bibinfo  {journal} {Nature Comm.}\ }\textbf {\bibinfo
  {volume} {9}},\ \bibinfo {pages} {768} (\bibinfo {year} {2018})}\BibitemShut
  {NoStop}%
\bibitem [{\citenamefont {van~der Linden}\ \emph {et~al.}(2019)\citenamefont
  {van~der Linden}, \citenamefont {Alexander}, \citenamefont {Aarts},\ and\
  \citenamefont {Dauchot}}]{PhysRevLett.123.098001}%
  \BibitemOpen
  \bibfield  {author} {\bibinfo {author} {\bibfnamefont {M.~N.}\ \bibnamefont
  {van~der Linden}}, \bibinfo {author} {\bibfnamefont {L.~C.}\ \bibnamefont
  {Alexander}}, \bibinfo {author} {\bibfnamefont {D.~G. A.~L.}\ \bibnamefont
  {Aarts}}, \ and\ \bibinfo {author} {\bibfnamefont {O.}~\bibnamefont
  {Dauchot}},\ }\href@noop {} {\bibfield  {journal} {\bibinfo  {journal} {Phys.
  Rev. Lett.}\ }\textbf {\bibinfo {volume} {123}},\ \bibinfo {pages} {098001}
  (\bibinfo {year} {2019})}\BibitemShut {NoStop}%
\bibitem [{\citenamefont {Quincke}(1896)}]{Quincke:96}%
  \BibitemOpen
  \bibfield  {author} {\bibinfo {author} {\bibfnamefont {G.}~\bibnamefont
  {Quincke}},\ }\href@noop {} {\bibfield  {journal} {\bibinfo  {journal} {Ann.
  Phys. Chem.}\ }\textbf {\bibinfo {volume} {59}},\ \bibinfo {pages} {417}
  (\bibinfo {year} {1896})}\BibitemShut {NoStop}%
\bibitem [{\citenamefont {Jones}(1984)}]{Jones:84}%
  \BibitemOpen
  \bibfield  {author} {\bibinfo {author} {\bibfnamefont {T.~B.}\ \bibnamefont
  {Jones}},\ }\href@noop {} {\bibfield  {journal} {\bibinfo  {journal} {IEEE
  Trans. Ind. Appl.}\ }\textbf {\bibinfo {volume} {4}},\ \bibinfo {pages} {845}
  (\bibinfo {year} {1984})}\BibitemShut {NoStop}%
\bibitem [{\citenamefont {Das}\ and\ \citenamefont
  {Saintillan}(2013)}]{Das:13}%
  \BibitemOpen
  \bibfield  {author} {\bibinfo {author} {\bibfnamefont {D.}~\bibnamefont
  {Das}}\ and\ \bibinfo {author} {\bibfnamefont {D.}~\bibnamefont
  {Saintillan}},\ }\href@noop {} {\bibfield  {journal} {\bibinfo  {journal}
  {Phy. Rev. E}\ }\textbf {\bibinfo {volume} {87}},\ \bibinfo {pages} {043014}
  (\bibinfo {year} {2013})}\BibitemShut {NoStop}%
\bibitem [{\citenamefont {Bricard}\ \emph {et~al.}(2013)\citenamefont
  {Bricard}, \citenamefont {Caussin}, \citenamefont {Desreumaux}, \citenamefont
  {Dauchot},\ and\ \citenamefont {Bartolo}}]{Bricard13}%
  \BibitemOpen
  \bibfield  {author} {\bibinfo {author} {\bibfnamefont {A.}~\bibnamefont
  {Bricard}}, \bibinfo {author} {\bibfnamefont {J.-B.}\ \bibnamefont
  {Caussin}}, \bibinfo {author} {\bibfnamefont {N.}~\bibnamefont {Desreumaux}},
  \bibinfo {author} {\bibfnamefont {O.}~\bibnamefont {Dauchot}}, \ and\
  \bibinfo {author} {\bibfnamefont {D.}~\bibnamefont {Bartolo}},\ }\href@noop
  {} {\bibfield  {journal} {\bibinfo  {journal} {Nature}\ }\textbf {\bibinfo
  {volume} {503}},\ \bibinfo {pages} {95} (\bibinfo {year} {2013})}\BibitemShut
  {NoStop}%
\bibitem [{\citenamefont {Bricard}\ \emph {et~al.}(2015)\citenamefont
  {Bricard}, \citenamefont {Caussin}, \citenamefont {Das}, \citenamefont
  {Savoie}, \citenamefont {Chikkadi}, \citenamefont {Shitara}, \citenamefont
  {Chepizhko}, \citenamefont {Peruani}, \citenamefont {Saintillan},\ and\
  \citenamefont {Bartolo}}]{Bricard15}%
  \BibitemOpen
  \bibfield  {author} {\bibinfo {author} {\bibfnamefont {A.}~\bibnamefont
  {Bricard}}, \bibinfo {author} {\bibfnamefont {J.-B.}\ \bibnamefont
  {Caussin}}, \bibinfo {author} {\bibfnamefont {D.}~\bibnamefont {Das}},
  \bibinfo {author} {\bibfnamefont {C.}~\bibnamefont {Savoie}}, \bibinfo
  {author} {\bibfnamefont {V.}~\bibnamefont {Chikkadi}}, \bibinfo {author}
  {\bibfnamefont {K.}~\bibnamefont {Shitara}}, \bibinfo {author} {\bibfnamefont
  {O.}~\bibnamefont {Chepizhko}}, \bibinfo {author} {\bibfnamefont
  {F.}~\bibnamefont {Peruani}}, \bibinfo {author} {\bibfnamefont
  {D.}~\bibnamefont {Saintillan}}, \ and\ \bibinfo {author} {\bibfnamefont
  {D.}~\bibnamefont {Bartolo}},\ }\href@noop {} {\bibfield  {journal} {\bibinfo
   {journal} {Nature Comm.}\ }\textbf {\bibinfo {volume} {6}},\ \bibinfo
  {pages} {7470} (\bibinfo {year} {2015})}\BibitemShut {NoStop}%
\bibitem [{\citenamefont {Karani}\ \emph {et~al.}(2019)\citenamefont {Karani},
  \citenamefont {Pradillo},\ and\ \citenamefont {Vlahovska}}]{Karani2019}%
  \BibitemOpen
  \bibfield  {author} {\bibinfo {author} {\bibfnamefont {H.}~\bibnamefont
  {Karani}}, \bibinfo {author} {\bibfnamefont {G.~E.}\ \bibnamefont
  {Pradillo}}, \ and\ \bibinfo {author} {\bibfnamefont {P.~M.}\ \bibnamefont
  {Vlahovska}},\ }\href@noop {} {\bibfield  {journal} {\bibinfo  {journal}
  {Phys. Rev. Lett.}\ }\textbf {\bibinfo {volume} {123}},\ \bibinfo {pages}
  {208002} (\bibinfo {year} {2019})}\BibitemShut {NoStop}%
\bibitem [{\citenamefont {Nourhani}\ \emph {et~al.}(2017)\citenamefont
  {Nourhani}, \citenamefont {Brown}, \citenamefont {Pletzer},\ and\
  \citenamefont {Gibbs}}]{nourhani2017}%
  \BibitemOpen
  \bibfield  {author} {\bibinfo {author} {\bibfnamefont {A.}~\bibnamefont
  {Nourhani}}, \bibinfo {author} {\bibfnamefont {D.}~\bibnamefont {Brown}},
  \bibinfo {author} {\bibfnamefont {N.}~\bibnamefont {Pletzer}}, \ and\
  \bibinfo {author} {\bibfnamefont {J.~G.}\ \bibnamefont {Gibbs}},\ }\href@noop
  {} {\bibfield  {journal} {\bibinfo  {journal} {Advanced Materials}\ }\textbf
  {\bibinfo {volume} {29}},\ \bibinfo {pages} {1703910} (\bibinfo {year}
  {2017})}\BibitemShut {NoStop}%
\bibitem [{\citenamefont {Dillmann}\ \emph {et~al.}(2013)\citenamefont
  {Dillmann}, \citenamefont {Maret},\ and\ \citenamefont
  {Keim}}]{dillmann2013two}%
  \BibitemOpen
  \bibfield  {author} {\bibinfo {author} {\bibfnamefont {P.}~\bibnamefont
  {Dillmann}}, \bibinfo {author} {\bibfnamefont {G.}~\bibnamefont {Maret}}, \
  and\ \bibinfo {author} {\bibfnamefont {P.}~\bibnamefont {Keim}},\ }\href@noop
  {} {\bibfield  {journal} {\bibinfo  {journal} {The European Physical Journal
  Special Topics}\ }\textbf {\bibinfo {volume} {222}},\ \bibinfo {pages} {2941}
  (\bibinfo {year} {2013})}\BibitemShut {NoStop}%
\bibitem [{\citenamefont {Ebert}\ \emph {et~al.}(2009)\citenamefont {Ebert},
  \citenamefont {Dillmann}, \citenamefont {Maret},\ and\ \citenamefont
  {Keim}}]{ebert2009experimental}%
  \BibitemOpen
  \bibfield  {author} {\bibinfo {author} {\bibfnamefont {F.}~\bibnamefont
  {Ebert}}, \bibinfo {author} {\bibfnamefont {P.}~\bibnamefont {Dillmann}},
  \bibinfo {author} {\bibfnamefont {G.}~\bibnamefont {Maret}}, \ and\ \bibinfo
  {author} {\bibfnamefont {P.}~\bibnamefont {Keim}},\ }\href@noop {} {\bibfield
   {journal} {\bibinfo  {journal} {Review of Scientific Instruments}\ }\textbf
  {\bibinfo {volume} {80}},\ \bibinfo {pages} {083902} (\bibinfo {year}
  {2009})}\BibitemShut {NoStop}%
\bibitem [{\citenamefont {Nourhani}\ and\ \citenamefont
  {Saintillan}(2021)}]{NourhaniSaintillan2021}%
  \BibitemOpen
  \bibfield  {author} {\bibinfo {author} {\bibfnamefont {A.}~\bibnamefont
  {Nourhani}}\ and\ \bibinfo {author} {\bibfnamefont {D.}~\bibnamefont
  {Saintillan}},\ }\href@noop {} {\bibfield  {journal} {\bibinfo  {journal}
  {Phys. Rev. E}\ }\textbf {\bibinfo {volume} {103}},\ \bibinfo {pages}
  {L040601} (\bibinfo {year} {2021})}\BibitemShut {NoStop}%
\end{thebibliography}
\end{document}